\documentclass[aps,pre,reprint,superscriptaddress]{revtex4-1} 

\usepackage{amssymb}
\usepackage{amsmath} 
\usepackage{dcolumn}   
\usepackage{bm}        
\usepackage{graphicx}
\usepackage{natbib}
\usepackage{appendix}
\usepackage{hyperref}

\begin{document}

\setlength{\parskip}{0.3\baselineskip}

\title{Transport on a Lattice with Dynamical Defects}

\author{Francesco Turci$^{\dagger}$}
\affiliation{Laboratoire Charles Coulomb, Universit\'e Montpellier 2 and CNRS, 34095 Montpellier, Cedex 5, France}
\affiliation{Theory of Soft Condensed Matter, Universit\'e du Luxembourg, L-1511 Luxembourg, Luxembourg }

\author{Andrea Parmeggiani}
\affiliation{Laboratoire Charles Coulomb, Universit\'e Montpellier 2 and CNRS, 34095 Montpellier, Cedex 5, France}
\affiliation{Laboratoire de Dynamique des Interactions Membranaires Normales et
Pathologiques, UMR 5235 CNRS, Universit\'e Montpellier 2 and Universit\'e Montpellier 1, 34095 Montpellier ,Cedex 5, France}

\author{Estelle Pitard}
\affiliation{Laboratoire Charles Coulomb, Universit\'e Montpellier 2 and CNRS, 34095 Montpellier, Cedex 5, France}

\author{M. Carmen Romano}
\affiliation{SUPA, Institute for Complex Systems and Mathematical Biology, King's College,
University of Aberdeen,  Aberdeen AB24 3UE, United Kingdom}
\affiliation{Institute of Medical Sciences, Foresterhill, University of Aberdeen, Aberdeen AB25 2ZD, United Kingdom}

\author{Luca Ciandrini$^{\dagger,}$}
\thanks{Present address: Laboratoire de Dynamique des Interactions Membranaires Normales et Pathologiques, UMR 5235, Universit\'e Montpellier 2, luca.ciandrini@gmail.com}

\thanks{\\$^\dagger$These authors contributed equally to this work.}
\affiliation{SUPA, Institute for Complex Systems and Mathematical Biology, King's College,
University of Aberdeen,  Aberdeen AB24 3UE, United Kingdom}

\begin{abstract}
Many transport processes in nature take place on substrates, often considered as unidimensional lanes. These unidimensional substrates are typically non static: Affected by a fluctuating environment, they can undergo conformational changes. 
This is particularly true in biological cells, where the state of the substrate is often coupled to the active motion of macromolecular complexes, such as motor proteins on microtubules or ribosomes on mRNAs, causing new interesting phenomena. 
Inspired by biological processes such as protein synthesis by ribosomes and motor protein transport, we introduce the concept of localized dynamical sites coupled to a driven lattice gas dynamics. We investigate the phenomenology of transport in the presence of dynamical defects and find a novel regime characterized by an intermittent current and subject to severe finite-size effects. Our results demonstrate the impact of the regulatory role of the dynamical defects in transport not only in biology but also in more general contexts.
\end{abstract}

\maketitle
%
\indent Transport in biology is often carried out on complex substrates of relative small sizes, such as microtubule filaments or mRNA strands~\cite{alberts_molecular_2008}. These substrates are not rigid and can either actively or passively change their conformation in time. Remarkably, the dynamical features of the substrates are mostly coupled and regulated by their interplay with the transport process. In the case of protein synthesis, for example, ribosomes can encounter folded regions of the mRNA strand that obstruct the path and interfere with their movement. Those local regions act like a dynamical switch that obstructs, when folded, or allows, when unfolded, the passage of ribosomes. Importantly, their folding-unfolding dynamics is strongly coupled to the presence of ribosomes: a secondary structure cannot re-fold until the ribosome has moved forward.\\ %
\indent More generally, the traffic dynamics is highly enriched due to the coupling between the state of the substrate and the particles and despite its primary importance to the physics of transport phenomena, this coupling has been usually neglected in the literature, with the exception of some particular cases related to microtubules dynamics~\cite{johann_length_2012, *reese_crowding_2011}. 
Importantly, substrate dynamics and their bottlenecks are indeed coupled with the flow of carriers not only in biology but also in many other systems, such as intelligent traffic lights~\cite{fouladvand_intelligent_2004} or pedestrian traffic~\cite{jelic_bottleneck_2012}. An extensive theory explaining the emergent effects of coupling between substrate and particle transport is therefore necessary.\\
\indent In this work we introduce and study transport on a lattice with dynamical defects, i.e., sites whose features are dynamically coupled to the presence of particles.  To do so we use an approach based on a paradigmatic model in non-equilibrium statistical physics, namely, the totally asymmetric simple exclusion process (TASEP)~\cite{macdonald_kinetics_1968, *schmittman_statistical_1995, *schutz_exactly_2001, *chou_non-equilibrium_2011}. 
We define as dynamical defects sites with two possible conformational states that depend on their particle occupancy. In the presence of dynamical defects, a new phenomenology appears: the current-density relation presents a plateau characterized by an intermittent flow of particles that cannot be estimated by standard mean-field arguments. Remarkably, in this regime we find large finite-size effects that strongly affect the transport characteristics.

\section{The Model}
The standard TASEP is a model of particles moving  unidirectionally on a one-dimensional discrete lattice with a fixed hopping rate $\gamma$. Steric interaction excludes that more than one particle can occupy the same site, and overtaking is not allowed. The average TASEP current of particles $J$ as a function of particle density $\rho$ is given by the parabolic relation $J_{TASEP}(\rho)=\gamma\rho(1-\rho)$ in the large lattice limit. The presence of one or more static defects in the lattice, defined as ``slow'' sites with hopping rates smaller than $\gamma$, leads to a reduction of the maximal current flow~\cite{janowsky_exact_1994} and confers a truncated parabola shape to the current-density relation $J(\rho)$. The defects considered so far in the literature are static since the hopping rates associated with the slow sites are constant. \\
\begin{figure}[tb]
\begin{center}
\includegraphics[width=0.82\columnwidth]{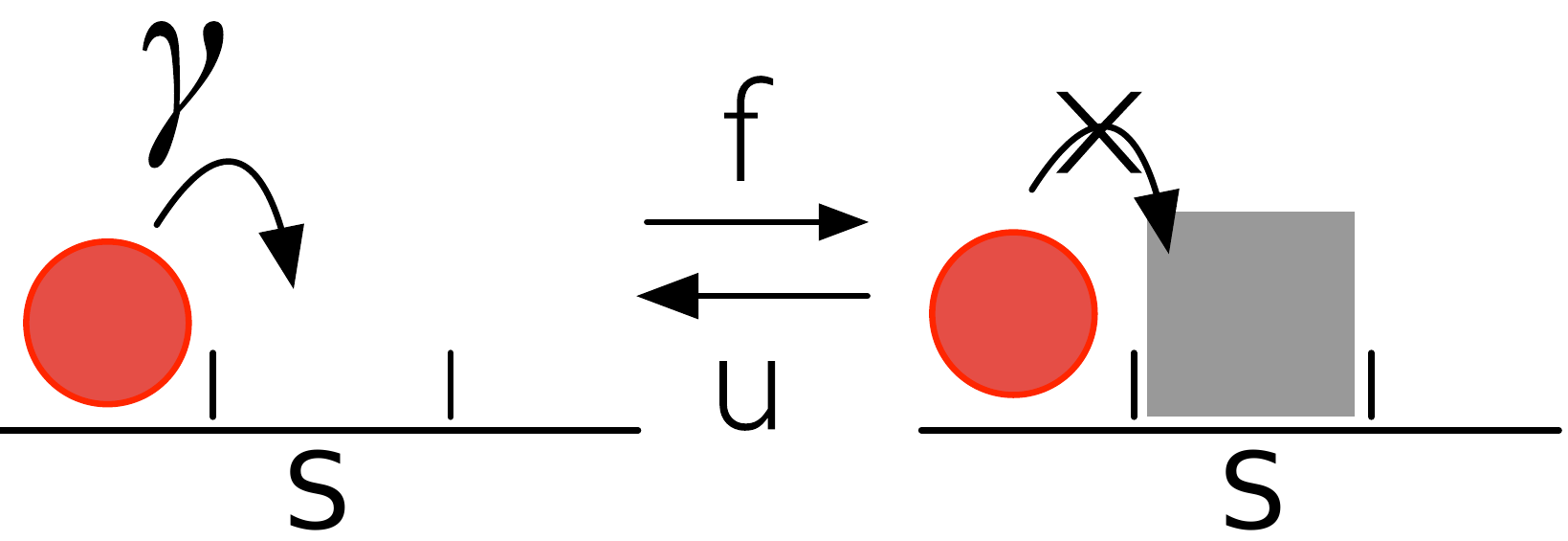}
\caption{(Color online) Mechanism for a one-site dynamical defect: the site $s$ closes or opens with rates $f$ and $u$, respectively,  represented by a gray square that obstructs the passage of particles. Particles hop at rate $\gamma$ and cannot enter the closed region. Moreover, if a particle occupies the site $s$, the closing of that site is forbidden.}
\label{TASEPrules}
\end{center}
\end{figure}

The periodic boundary case illustrates well the main phenomenology of the model. We consider therefore a ring of $L$ sites on which $N \le L$ particles are allowed to hop in one direction, fixing the overall density at $\rho = N/L$. 
A region of size $d$ represents the dynamical defect and has an intrinsic two-state dynamics coupled to the presence of particles.
The defect can pass from the \textit{closed} state to the \textit{open} state with rate $u$. The inverse process occurs with rate $f$ only when all $d$ sites of the defect are empty.
Our results indicate that the single-site defect case for $d=1$ (see Fig.~\ref{TASEPrules}) exhibits the main features of the model. Therefore in this work we focus, for sake of simplicity, on this case. The model proposed here can be applied to describe further systems; for instance, junctions in transport networks~\cite{embley_understanding_2009} can be thought of as particular dynamical defects with influx dependent dynamics.

\section{Numerical Simulation Results}
\begin{figure}[t!]
\begin{center}
	\includegraphics[width=\columnwidth]{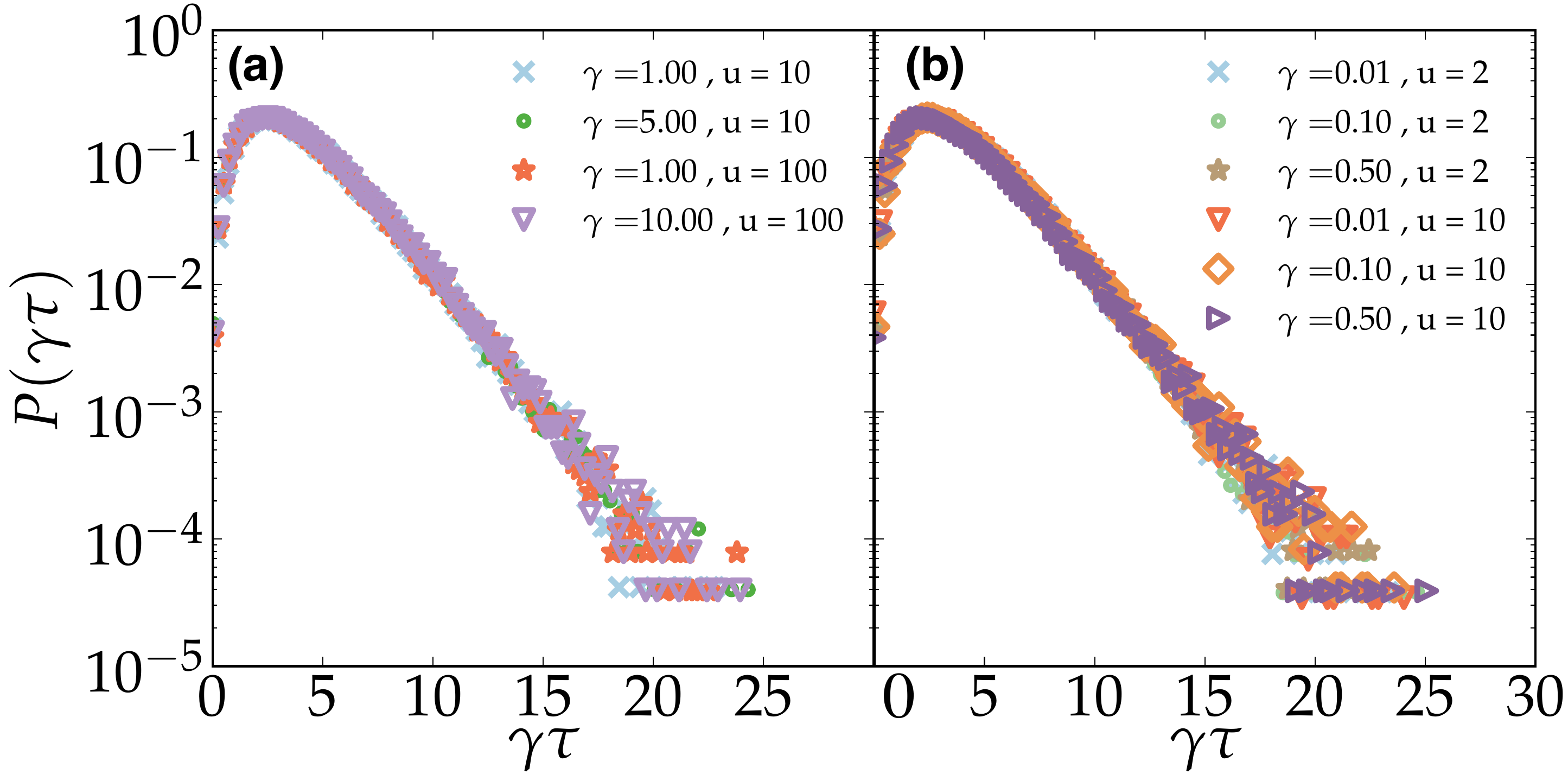}
\caption{(Color online) Time lag distributions rescaled by $\gamma$ in the TASEP-like limit for a system of size $L=500$ at density $\rho=0.5$ for (a) $u > \gamma > f$ and (b) $u > f > \gamma$ ($f=1$ in both cases).}
\label{fig:s1}
\end{center}
\end{figure}

We have performed continuous-time Monte Carlo simulations based on the Gillespie algorithm~\cite{gillespie_general_1976}. We have studied the model in a large parameter space taking several lattice sizes $L$ (up to $4000$) and varying the rates $\gamma, u, f$ in order to explore all the different dynamical regimes of the model for which we determined the characteristic timescales. 
We numerically characterized the different regimes by computing the probability distribution function of the time lags $\tau$ between the passage of two consecutive particles on one site. In particular, we chose the site $s+1$, right after the defect site $s$. \\
When the opening rate $u$ is the largest rate (i.e. when $u > f > \gamma$ or $u > \gamma > f$)  we find a single timescale governed by the hopping rate $\gamma$. Particles essentially flow without a significant interaction with the defect.
Figure~\ref{fig:s1}, indeed, shows that the time lag distributions collapse if rescaled by the hopping rate $\gamma$. We naturally define such a behavior as the TASEP-like regime.

\begin{figure}[t!]
\begin{center}
	\includegraphics[width=0.8\columnwidth]{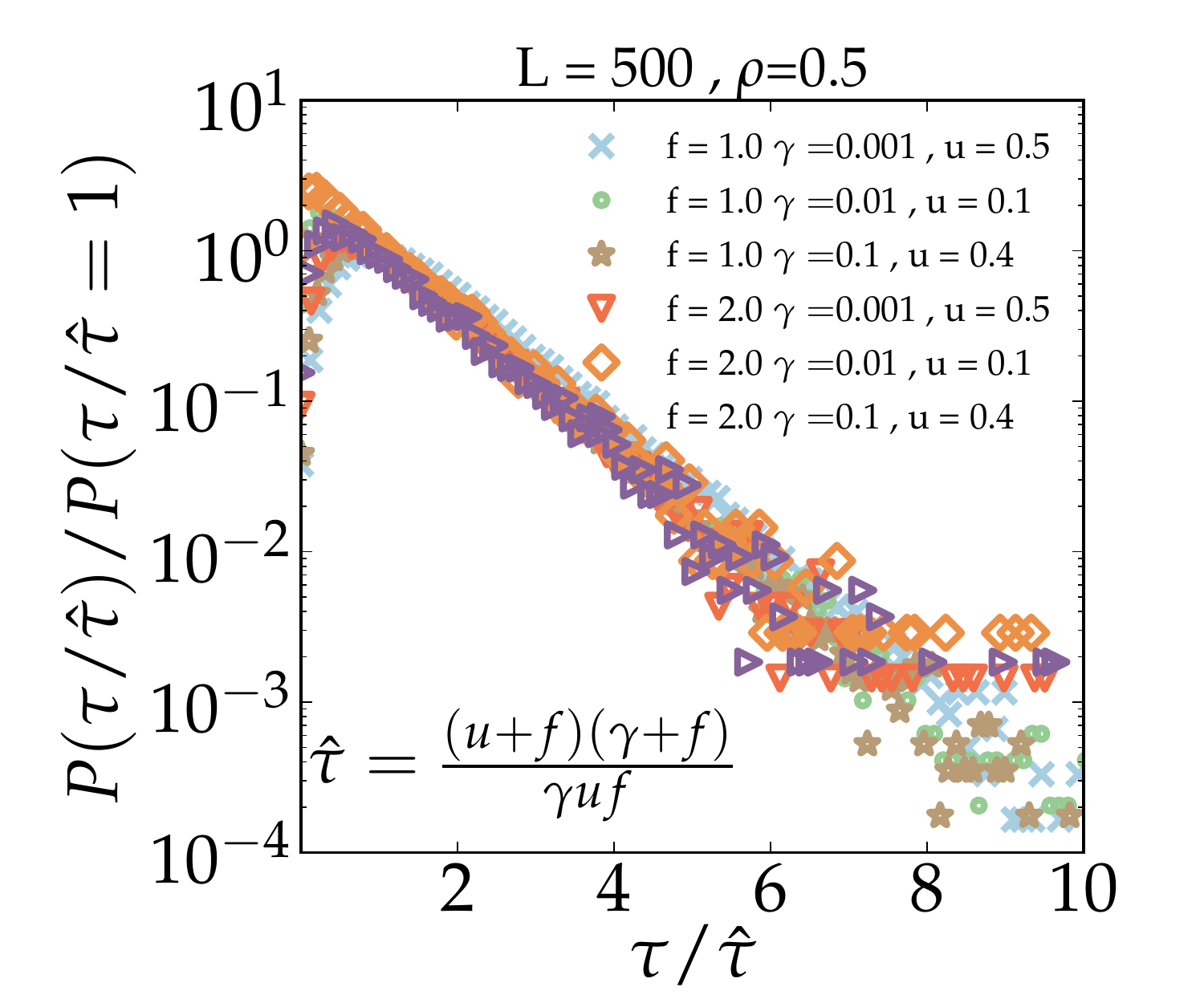}
\caption{(Color online) Time lag distributions in the slow site regime $f>u> \gamma$ rescaled by $\hat{\tau}$, see Eq.~(\ref{tau}). \label{fig:s2}}
\end{center}
\end{figure}

Similarly,  when the closing rate $f$ is the largest rate and particles can pass only one at a time during an opening event (i.e. when  $f>u>\gamma$), the time distribution is characterized by a single characteristic timescale for long times.\\
In such a regime the dynamical defect acts like a static defect.
One can therefore think that particles are injected into the region after the defect with an effective constant entry rate $q$, in analogy to a slow site or static defect~\cite{Janowsky:1992p15963}. Such an effective rate can be approximated by the product of the hopping rate and the probability that the defect region is open:
\begin{equation}
q=\gamma \frac{u}{u+f} \;.
\end{equation}
We also note that the probability to move a particle onto the defect, when this is open, is $\gamma/(\gamma+f)$.\\
\indent If we denote by $\rho_{i}^{open}$ the probability to find a particle on site $i$ restricted to times when the defect is open, the current can be approximated by $J=q \rho_{s-1}^{open}(1-\rho_{s}^{open})$ where $s$ is the defect site. The typical passage time will scale then as $\hat{\tau}\approx 1/J$, which is the typical time for a particle to pass.  By assuming, due to the blockage, that $\rho_{s-1}^{open}\approx 1$, the probability to find a particle on site $s$ when the region is open can be approximated by the probability to move a particle there, i.e. $\rho_{s}^{open}\sim\gamma/(\gamma+f)$.\\
This leads to an estimate of the typical timescale $\hat{\tau}$ in this regime
\begin{equation}
	\hat{\tau}\approx \frac{(u+f)(\gamma+f)}{\gamma u f }\;,
	\label{tau}
\end{equation}
which is in good agreement with the simulations, see Figure~\ref{fig:s2}.

\begin{figure}[!t]
\begin{center}
	\includegraphics[width=0.9\columnwidth]{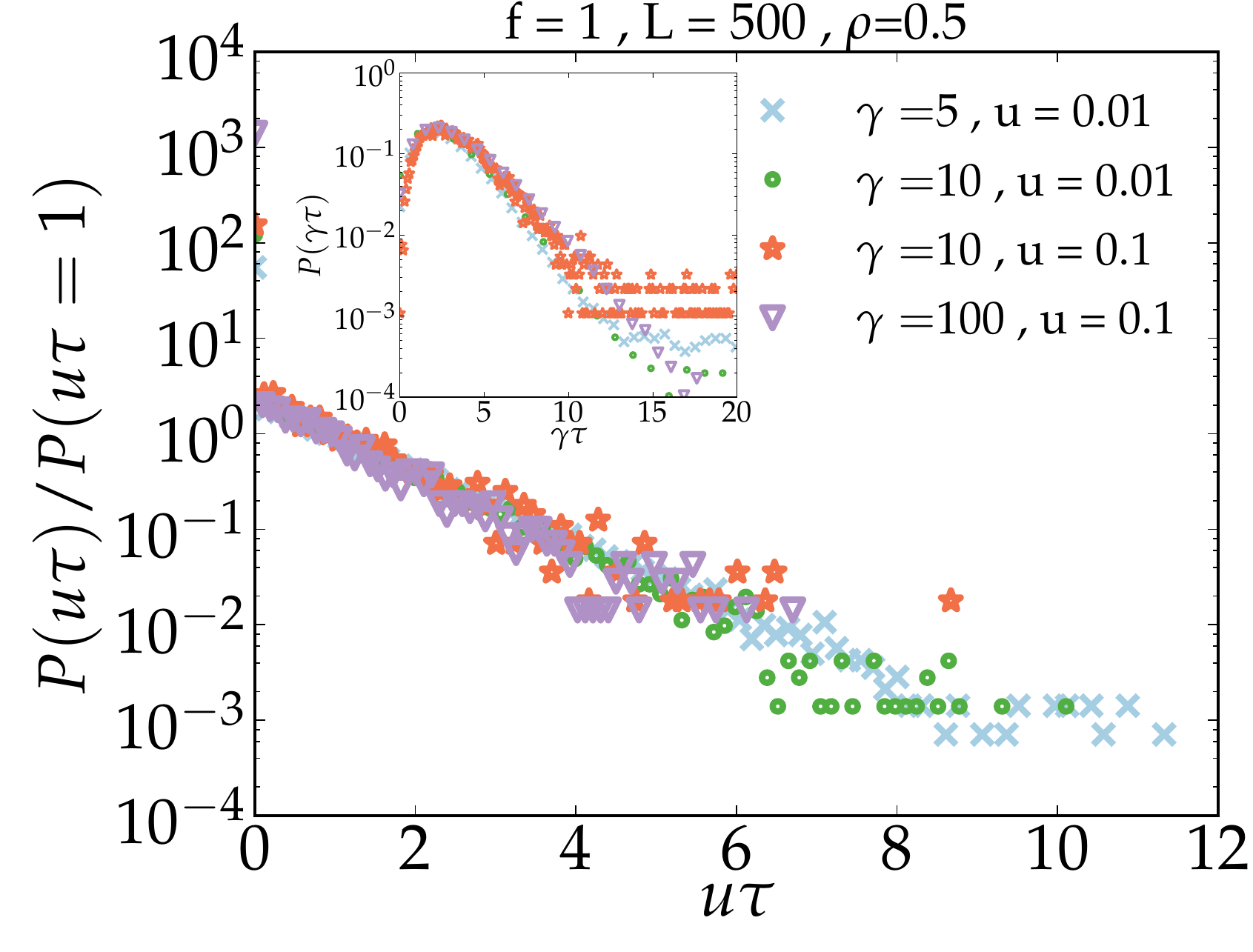}
\caption{(Color online) Time lag distributions in the intermittent regime $\gamma>f>u$. The short time dynamics is governed by the hopping rate $\gamma$ (see inset) while the long time by the opening rate $u$.}
\label{fig:s3}
\end{center}
\end{figure}

The situation changes in the other regimes, and in particular when $\gamma>f>u$. Here two timescales are present, consisting in a sharp peak and a large tail in the time lag distributions (see Fig.~\ref{fig:s3}). The sharp peak is a signature of several particles passing through the blockage during the same opening event, while the large tails are given by the long waiting times to open the region (since $u$ is the smallest rate). 

We therefore expect that the short time dynamics is regulated by the open, TASEP-like behavior of the system, and therefore described by the rate $1/\gamma$ , while the long time dynamics is governed by $1/u$. This is consistent with the outcome of the simulations, as shown in Figure~\ref{fig:s3}. Two regimes are clearly distinguishable: in the first one  the dynamics is the same as in the TASEP and in the second one the distributions of time lags shows an exponential tail. We call this regime the intermittent regime.

\begin{figure}[h!tbp]
\begin{center}
	\includegraphics[width=0.95\columnwidth]{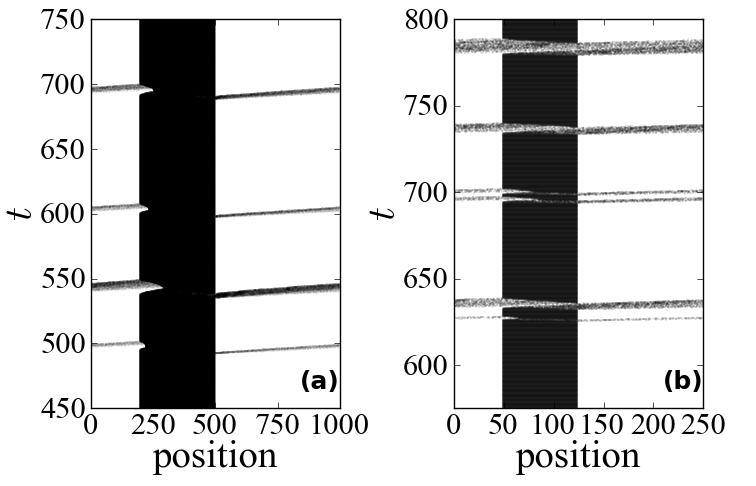}
\caption{Kymographs of a system in the intermittent regime ($\rho=0.3$, $u=0.01$, $f=1$, and $\gamma=100$) for (a) a large lattice $L=1000$ and (b) a small lattice $L=250$. The defect is located in the middle of the lattice. \label{fig:s5}}
\end{center}
\end{figure}

The intermittent regime can be easily visualized by the use of so-called \textit{kymographs}, space-time representations of the evolution of the system. Figure~\ref{fig:s5}a shows that, for large sizes, there are always high densities (HD) and low density (LD) regions before and after the defect, respectively. The site is often closed and does not allow the flow of particles; when it opens, several particles are able to pass. This creates an intermittent behavior of the current.

In the case of small systems, as depicted in Figure~\ref{fig:s5}b for a system of $L=250$, the unstable HD-LD front relaxes over the entire lattice length. This causes severe finite-size effects, which will be discussed in Section~\ref{finite-size}.

\begin{figure*}[t!]
\begin{center}
\includegraphics[width=0.72\textwidth]{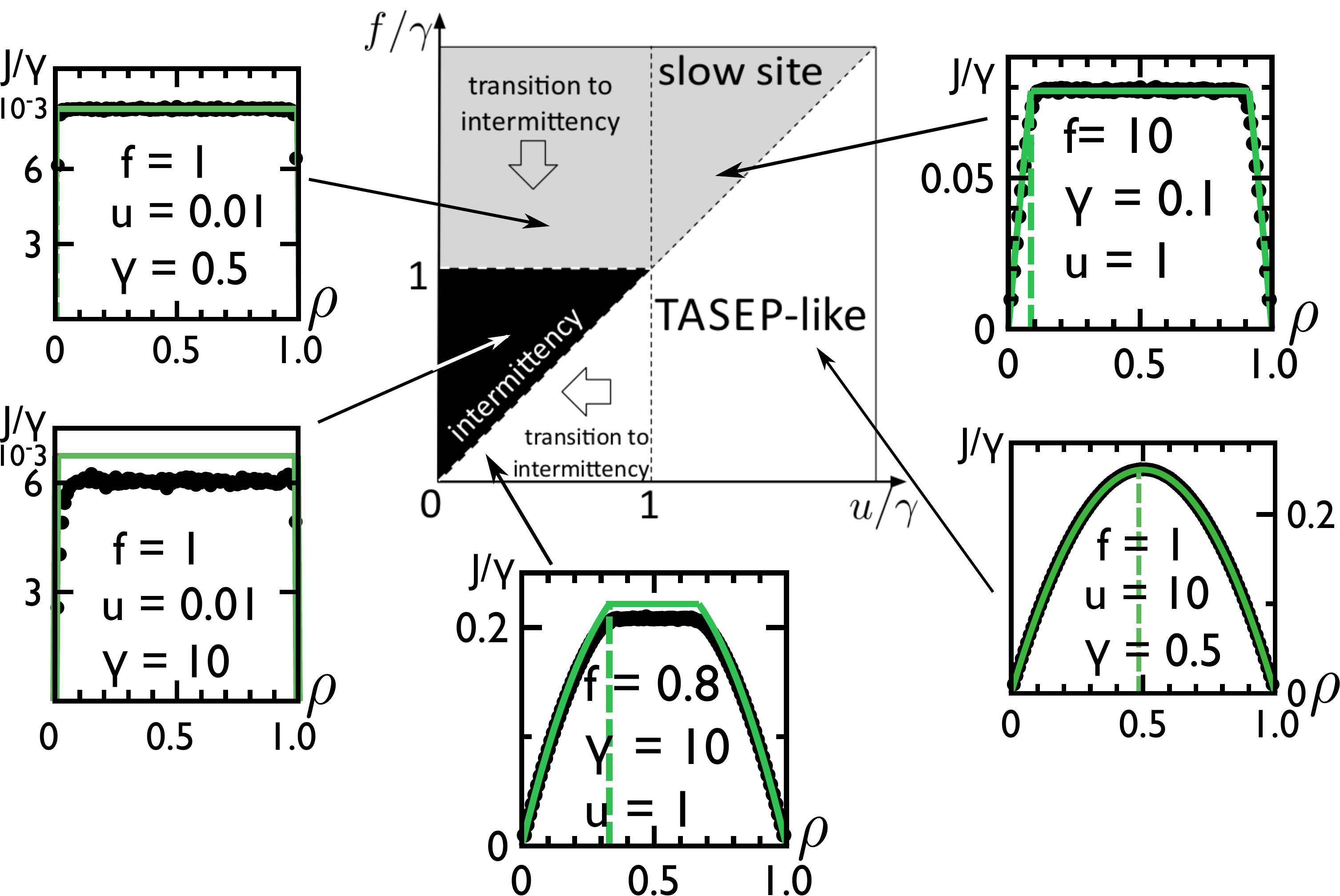}
\caption{(Color online) Sketch of the phase diagram of the system in the $f/\gamma$ and $u/\gamma$ space. The different regimes are separated by dashed lines. The atypical current-density relations are shown for every regime: numerical simulations (closed circles) are compared with the finite-segment mean-field (FSMF) predictions (continuous lines) introduced in Sec.~\ref{sec::FSMF}. While away from the intermittent regime the agreement is satisfactory, deep in the intermittent (black) regime the FSMF prediction overestimates the current.}
\label{folding-diag-composite}
\end{center}
\end{figure*}

The phase diagram shown in Fig.~\ref{folding-diag-composite} summarizes the behavior of the system described so far. Each regime exhibits a different slope in a log-linear scale defining characteristic timescales that we sum up as it follows:
\begin{itemize}
\item[(i)]{\it The TASEP-like regime}: when $u>f>\gamma$ or $u>\gamma>f$, due to the rapid opening, the current is not affected by the defect. The system therefore behaves like a homogeneous TASEP and the only relevant timescale is $1/\gamma$ (white region in Fig.~\ref{folding-diag-composite});
\item[(ii)]{\it The static defectlike regime}: when $f>u>\gamma$ the dynamical defect acts like a static one, allowing the passage of  particles in a way that can be described with a single effective rate $q<\gamma$ (top-right gray region in Fig.~\ref{folding-diag-composite}), similar to the model in~\cite{janowsky_exact_1994};
\item[(iii)]{\it The intermittent regime}: when $\gamma>f>u$ there are two strongly separated timescales, in contrast to the former cases. The short timescale $1/\gamma$ is the time separation between the passage of two consecutive particles during an opening event, while the long timescale $1/u$ corresponds to the time intervals during which the defect is closed and obstructs the passage of particles (black region in Fig.~\ref{folding-diag-composite}). This is a different regime caused by the presence of the dynamical defect.
\end{itemize}
The remaining cases ($\gamma>u>f$ and $f>\gamma>u$) are crossovers regimes between the intermittent and the homogeneous or static-defect TASEP-like behaviors, respectively.

These different regimes have a specific signature in the current versus density relation $J(\rho)$.
Similarly to the cases presented in the literature involving localized static defects in one-dimensional systems~\cite{janowsky_exact_1994}, numerical simulations show a reduction of the current with respect to the homogeneous TASEP; the current-density relation is a truncated parabola with a constant plateau value, see insets in Fig.~\ref{folding-diag-composite}. In large systems, when the opening rate $u$ decreases with respect to the other typical timescales, the current-density plateau lowers to smaller values, occupying a wider interval of densities and merging with the TASEP parabola $\gamma\rho(1-\rho)$ only for $\rho \to 0$ and $\rho \to 1$. As in~\cite{janowsky_exact_1994}, clusters of particles form before the defect so that the average density profiles are usually characterized by a sharp separation between a HD and a LD phase. In regions of the parameter space where the dynamics of the defect does not play a major role, our model recovers the standard phenomenology of an homogeneous exclusion process and  the one of an exclusion process with a static-defect. However, we shall show that when the dynamics of the defect induces the intermittent regime, the system is subject to severe finite-size effects which considerably modify the current versus density relation $J(\rho)$ and the average density profiles in space $\rho(x)$. We will present and discuss these finite-size effects in Section \ref{finite-size}.
\section{Mean-field approaches}
\subsection{Finite-segment mean-field}
\label{sec::FSMF} 
To compute the current-density relationship we use a finite-segment mean-field (FSMF) approach~\cite{*chou_clustered_2004, *dong_understanding_2009}, which allows us to define effective entry and exit rates from the pair of sites ($s-1,s$), where the dynamics is treated exactly.
%

We illustrate the FSMS approximation as it follows. We consider a ring of $L$ sites and a dynamical defect composed of one single site at position $s$. In the large-$L$ limit we can imagine splitting the system into three parts: a semi-infinite left sub-lattice, a semi-infinite right sub-lattice and in between a middle region composed of the defect at sites $s$ and $s-1$. We study then the dynamics in the middle region introducing effective rates for the injection and extraction of particles. 
Denoting the dynamical defect site by $s$, the pair of sites ($s-1,s$) has six possible states, given that site $s-1$ can be occupied or empty, and site $s$ can be empty and open, empty and closed, or occupied and open (see Table~\ref{table_states}).
\begin{table}[h!]
\begin{center}
\begin{tabular}{l|l|l|l}
label & s-1 & s & conformation of $s$\\
\hline
$x_{1}$ & 0 & 0 & open\\
$x_{2}$  & 0 & 0 & closed\\
$x_{3}$ & 1 & 0 & open\\
$x_{4}$ & 1 & 0 & closed\\
$x_{5}$ & 0 & 1 & open\\
$x_{6}$ & 1 &1 & open
\end{tabular}
\end{center}
\caption{Available configurations of sites $s-1$ and $s$. \label{table_states}}
\end{table}\\
\indent When the current reaches its plateau the system is split into HD and LD phases, separated by the dynamical defect. By imposing then current continuity, the two densities have to be coupled as $\rho_{HD}=1-\rho_{LD}=1-\rho_{s}$ ($\rho_s$ denotes the density of the defect site and we assume that site $s$ is in the LD phase).\\
The master equation 
\begin{equation}
\label{eq:me}
\frac{\partial \vec{P}}{\partial t}=\mathbb{W}\vec{P}
\end{equation}
for the probability $\vec{P}$ to find the system in one of the six states $x_1, x_2,\ldots x_6$ is well defined once all the transition rates between all different states are specified. The transition matrix $\mathbb{W}$ between the states  then reads
\begin{equation}
\small
\mathbb{W}=\left(
\begin{array}{cccccc}
 -f-\gamma  \hat{\rho}_{s} & u & 0 & 0 & \gamma \hat{\rho}_{s} & 0 \\
 f & -u-\gamma  \hat{\rho}_{s} & 0 & 0 & 0 & 0 \\
 \gamma  \hat{\rho}_{s} & 0 & -f-\gamma  & u & 0 & \gamma  \hat{\rho}_{s} \\
 0 & \gamma \hat{\rho}_{s} & f & -u & 0 & 0 \\
 0 & 0 & \gamma  & 0 & -2 \gamma  \hat{\rho}_{s} & 0 \\
 0 & 0 & 0 & 0 & \gamma \hat{\rho}_{s} & -\gamma  \hat{\rho}_{s}
\end{array}
\right) \;,
\end{equation}
\normalsize
where we used the notation $\hat{\rho}_{s}=1-\rho_{s}$.  The matrix $\mathbb{W}$ therefore contains the effective transition rates as a function of the density on the defect $\rho_{s}$, assuming that a shock is located in the middle region between a phase at high density $1-\rho_{s}$ and a phase at low density $\rho_{s}$. 

Solving the master equation (\ref{eq:me}) in the steady-state, we compute, as a function of $\rho_{s}, f, u$ and $\gamma$, the probability to find a particle on site $s$, which is by definition equal to the density $\rho_s$. One then obtains an expression for $\rho_{s}$ as a function of all other parameters (see the Appendix for more details).
The plateau current is then given by $J_{plateau}= \gamma \rho_{s} (1-\rho_{s})$. Note that this procedure can be extended to larger defects ($d>1$).

\begin{figure}[htb]
\includegraphics[width=0.45 \textwidth]{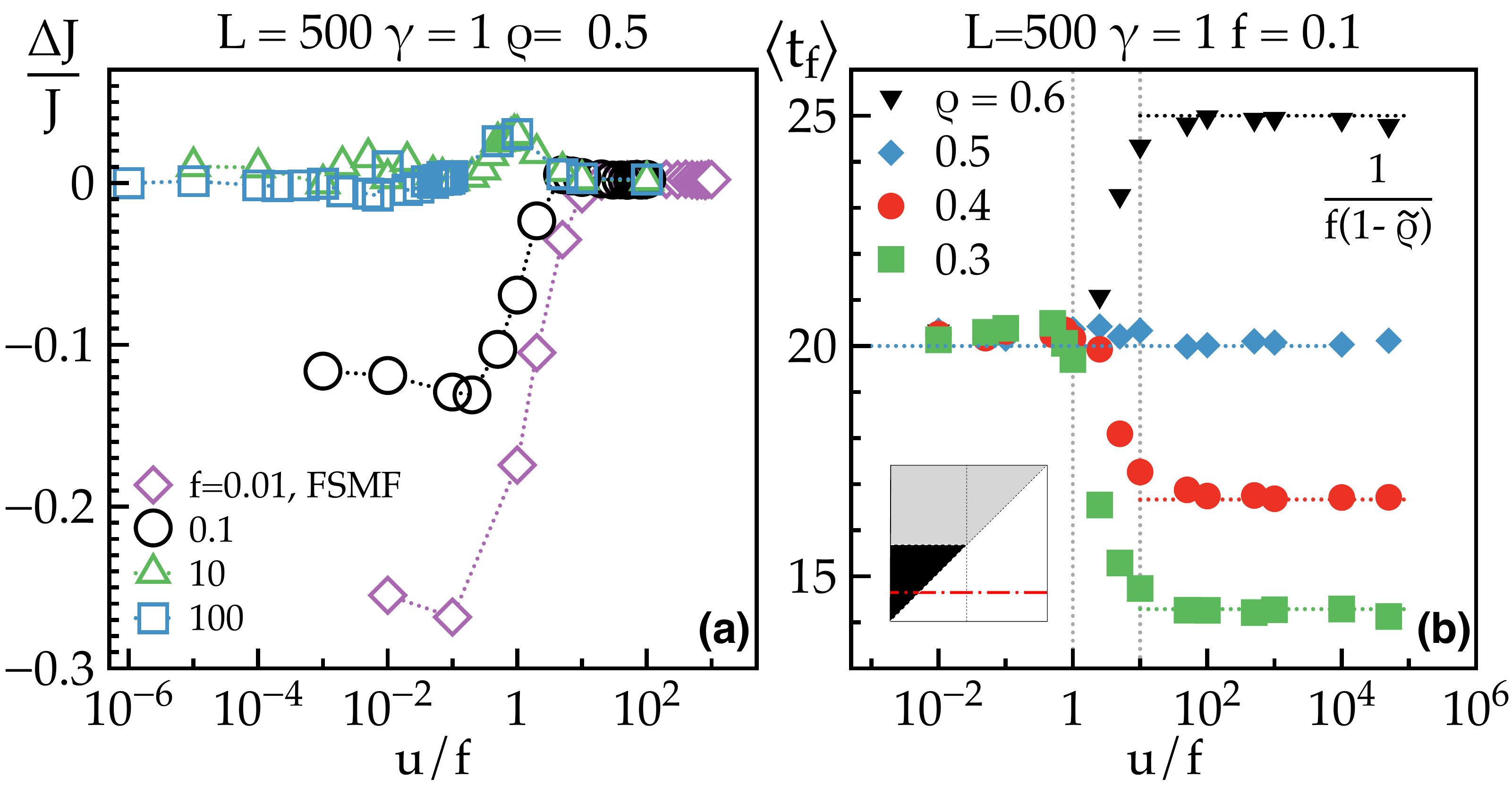} \hspace{-5ex}
\caption{(Color online) (a) Relative difference $\Delta J/J=(J-J_{FSMF})/J$ between simulations and the FSMF approach. Circles and diamonds show the crossover from  the intermittent regime to the TASEP-like regime, while squares and triangles show the crossover  from an intermediate nonintermittent regime to the slow-site-like regime, see Fig.~\ref{folding-diag-composite}. (b) Average closing times $\langle t_f\rangle$ (in seconds) for different densities. Vertical dotted lines represent the boundaries between the different regimes crossed (see, dash-dotted line in the inset). For large $u/f$, $\tilde{\rho}\sim \rho$, while in the intermittent region at small $u/f$ the assumption $\tilde{\rho}\sim 0.5$ provides a good estimate of the closing times. }
\label{fig::FSMF}
\end{figure}

To validate the FSMF approach, in Fig.~\ref{fig::FSMF}a we show the relative difference $\Delta J/J=(J-J_{FSMF})/J$ between simulations and the FSMF in all the different regimes of the phase diagram. Data are taken in the middle of the plateau of the current-density relation ($\rho=0.5$), also to avoid deviations due to finite-size effects (section  \ref{finite-size}). This analysis provides 
reasonably good results in the TASEP-like and the slow-site-like regimes. However, it reveals also that the FSMF approximation is not appropriate in the intermittent regime (circles and diamonds in Fig.~\ref{fig::FSMF}a, $u/f<1$).

\subsection{Intermittent mean-field}
\label{sec::intermittent}
We therefore turn our attention to the intermittent regime, where the FSMF approach fails. We start by analyzing the average time between consecutive opening and closing events $\langle t_{f} \rangle$, Fig.~\ref{fig::FSMF}b, i.e. the average time the folding region remains open allowing the passage of particles, before folding again.

When the coupling between the conformation of the defect and the presence of particles is weak, i.e. $u$ is the largest rate, the average timescales as $\langle t_{f}\rangle \sim f^{-1}$. However, when the coupling is strong, i.e., $\gamma$ is the largest rate, the former expression is corrected as $\langle t_{f}\rangle=[f(1-\tilde{\rho})]^{-1}$, where $\tilde{\rho}$ is the probability to find a particle on the defect site given that the site $s$ is open (previously also called $\rho_s^{open}$). This probability can be approximated in some limiting cases: in the TASEP-like regime, $u$ is the fastest rate and hence the defect is almost always open. In this case $\tilde{\rho}\sim \rho$ and the current is very close to the one predicted by the pure TASEP. 

In contrast, in the intermittent regime $J(\rho)$ exhibits a plateau (see the insets in Fig.~\ref{folding-diag-composite}), within which the opening-closing dynamics is independent of the total density. In this case particles are blocked for long times behind the closed defect that occasionally opens, thus allowing for a collective passage of particles, as confirmed by the kymographs (see Fig.~\ref{fig:s5}). Just after the opening of the defect, $\tilde{\rho}$ can be estimated from a simple mean-field approach: $d\tilde{\rho}/dt=\gamma \rho_{s-1}^{open} (1-\tilde{\rho})-\gamma\tilde{\rho}(1-\rho_{s+1}^{open})$, where $\rho_{s-1}^{open}$ and $\rho_{s+1}^{open}$ denote the density of particles before and after the defect, respectively. Approximating $\rho_{s-1}^{open}$ by 1 and $\rho_{s+1}^{open}$ by 0, one gets $\tilde{\rho}=0.5$ in accordance with numerical simulations, as Fig.~\ref{fig::FSMF}b shows a good agreement with the folding times $\langle t_f \rangle$ in the intermittent regime estimated with $\tilde{\rho}=0.5$. \\
%
\indent Hence, an intermittent dynamics is established so that no net flow passes through the defect for a time of order $1/u$ and then a large current $J_{open}=\gamma \tilde{\rho}(1-\tilde{\rho})$ flows for a time $\langle t_{f}\rangle$.
This gives an average plateau current
 \begin{equation}
J_{IMF}=\gamma \tilde{\rho}(1-\tilde{\rho})\dfrac{u/f}{1-\tilde{\rho}+u/f}
\label{eq:iMF}
\end{equation}
which we refer to as intermittent mean-field (IMF) current.
Figure~\ref{phase}a shows that deep in the intermittent regime (where $\tilde\rho\sim 0.5$) this approach provides a substantial improvement over the FSMF approach in predicting the current in the plateau.

\section{Finite-size Effects}
\label{finite-size}
\begin{figure}[htb]
\includegraphics[width=\columnwidth]{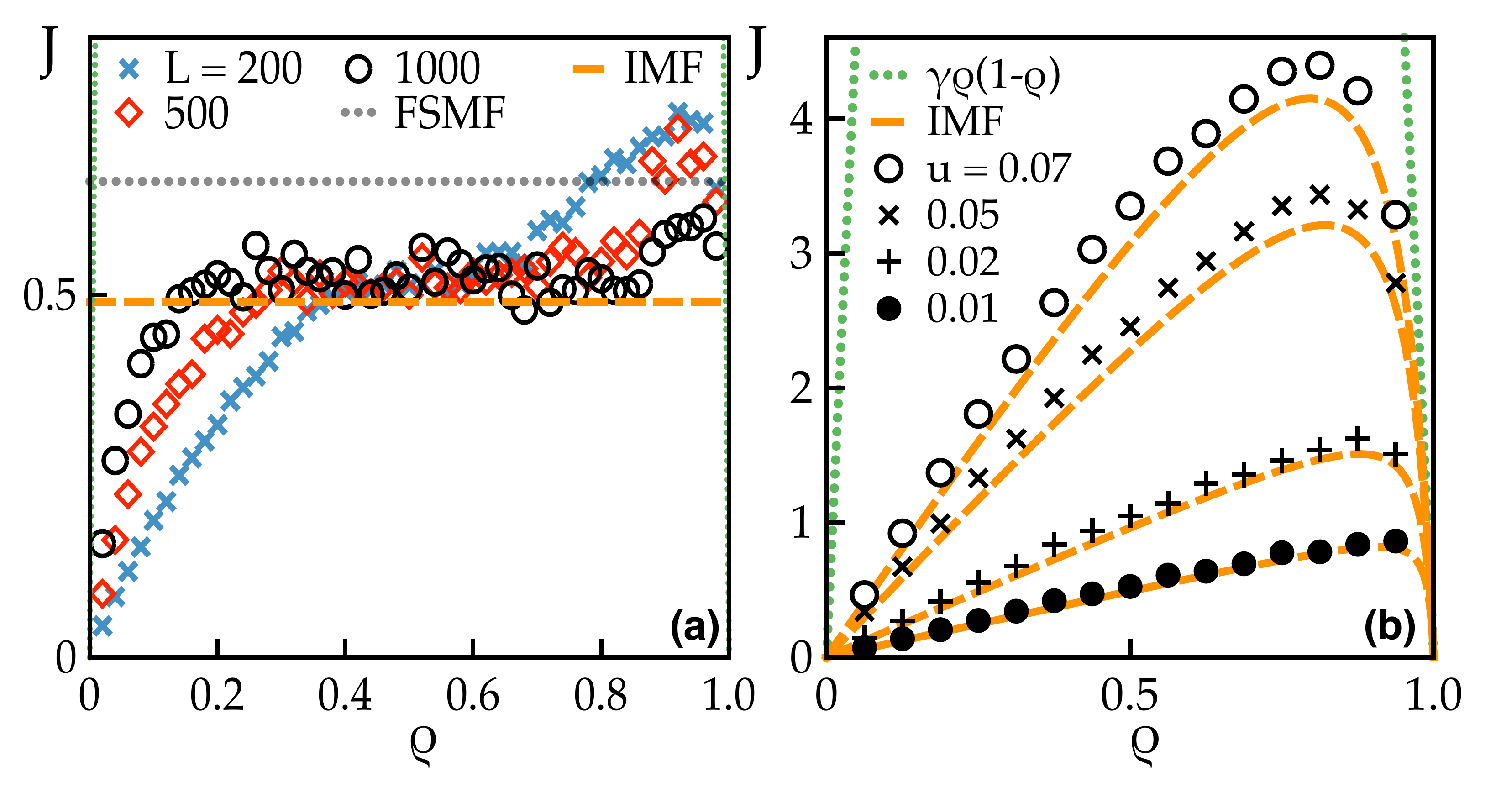}
\caption{(Color online) (a) Deep in the intermittent regime the FSMF calculation (dotted line) is unable to predict the plateau value, while the IMF with $\tilde{\rho}=0.5$ (dashed line) does. Moreover, in the intermittent regime truncated parabolas are found only in the large-$L$ limit: smaller systems show current reduction (enhancement) at low (high) densities. The parameter used are $f=1s^{-1}, u=0.01s^{-1}$ and $\gamma=100s^{-1}$. (b) For very small systems (here $L=16$) the IMF calculation with $\tilde{\rho}=\rho$ (dashed lines) correctly approximates the numerical $J(\rho)$ relation ($f=1s^{-1}, \gamma=100s^{-1}$).}
\label{phase}
\end{figure}

As previously stated, the system presents strong finite-size effects, remarkably pronounced in the intermittent regime. Here the current-density relation of small lattices becomes asymmetric (Fig.~\ref{phase}a): the current is reduced at small densities whereas it is enhanced for large densities. Figure~\ref{phase}a also shows that the value of $J$ in the plateau does not depend on the system size. Moreover, the plateau disappears for very small systems (Fig.~\ref{phase}b) while the current-density profile remains asymmetric.
    
Such a behavior can be rationalized as follows: whereas in the thermodynamic limit the probability to find a particle on the defect site, when open, is well approximated by $\tilde{\rho}\sim 0.5$, in very small systems (e.g. $L=16$ in Fig.~\ref{phase}b) the unstable HD-LD interface moving from the dynamical defect quickly relaxes over the whole system to the homogeneous TASEP density. Therefore, when the defect allows the passage of particles, all the sites can be considered to be identical and the density on the defect $\tilde \rho \sim \rho$. 
Thus, identifying $\tilde{\rho}$ with $\rho$ in Eq.~(\ref{eq:iMF}) gives good quantitative agreement, shown in Fig.~\ref{phase}b. Moreover, this reasoning allows the determination of the lower boundary for the current versus density relation $J_{L}(\rho)$ for very small system sizes $L$, the upper boundary $J_{\infty}(\rho)$ being the truncated parabola profile predicted by the IMF theory.

For intermediate system sizes $L$ the situation is again different. Remarkably, the current for $\rho>0.5$ is enhanced compared to the current obtained in large systems, whereas it is reduced for $\rho<0.5$ (Fig.~\ref{phase}a).  
This effect can be understood by analyzing the relaxation dynamics after an opening event by integrating the system of $L$ coupled ordinary
differential equations (ODEs) describing the occupancy of each site in the lattice $\rho_i$.
In this respect, we examine the relaxation of a TASEP system with initial conditions representing particles queuing behind the closed region and waiting for its opening. 
Before the opening, a HD region forms before the defect site $s$, while starting from the same defect site the system is at LD.
Therefore we want to study the unstable HD-LD front moving through the folding site during the interval of time between two folding events. Like the situation represented in Fig.~\ref{fig:s5}, we imagine starting with $N=L\rho$ particles queued behind the closed site $s$. 
At time $t=0$ the site opens, letting the particles flow and relaxing the HD-LD inhomogeneity. We focus then on how the density $\tilde\rho$ that a particle occupies the site $s$ (impeding its closing) evolves in time between two closing events. In order to do so, we
 write a system of $L$ coupled differential equations
\begin{equation}
	\frac{d\rho_i}{dt} = \gamma \rho_{i-1} (1-\rho_i) - \gamma \rho_{i} (1-\rho_{i+1}), \qquad i=1,\dots,L \,,
	\label{eq:ODE}
\end{equation}
with the prescription $\rho_0\equiv \rho_L$ (periodic boundary conditions), and fixing the initial conditions as described above, i.e. $\rho_i=0$, $\forall \, i$ excluding the $N$ sites preceding the defect for which $\rho_{s-j}=1$, $j=1,\dots N$, where $s$ is the defect site and $N= L \rho$ is the total number of particles (here $\gamma$ is arbitrarily fixed to 1 and defines the unit of time).\\
\indent We then imagine opening the dynamical defect and integrating the system numerically to observe the evolution, with time, of the occupancy $\tilde{\rho}$ of the site $s$ after the opening event. Even though here we do not consider successive closing events (we are only interested in the relaxation dynamics), in the intermittent regime we consider that, before any opening of the site $s$, the system lies in a situation very well approximated by the previous initial conditions, as supported also by the kymographs in Fig.~\ref{fig:s5}.

\begin{figure*}[!th]
\begin{center}
	\includegraphics[width=0.7\textwidth]{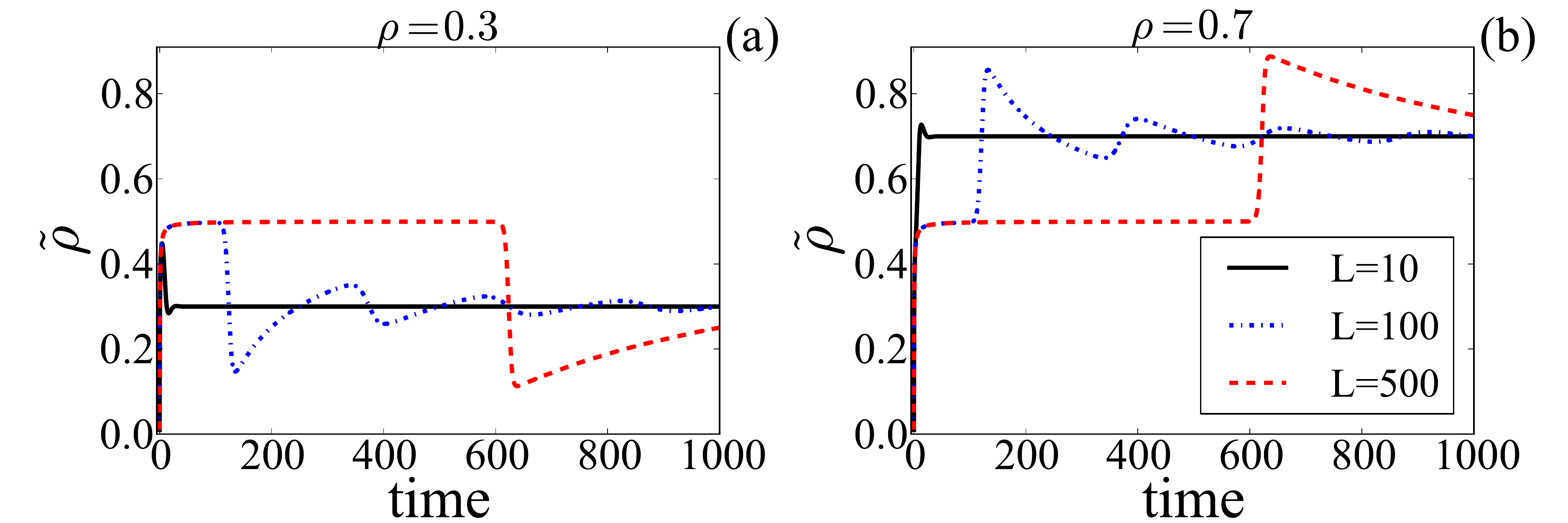}
\caption{(Color online) Numerical solution of $\tilde \rho (t)$ based on the system of ODEs~(\ref{eq:ODE}). The black curve shows that the steady state $\tilde \rho = \rho$ is reached quickly in small systems, and through an oscillatory transient in larger systems (dashed and dash-dotted lines). \label{fig:s6}}
\end{center}
\end{figure*}

The results, shown in Fig.~\ref{fig:s6}, strongly depend on the size $L$ of the system. The density $\tilde{\rho}$ of small systems (with just a few sites) relaxes very quickly to the homogeneous TASEP density. This is the reason why, for very short lattices, we can approximate $\tilde{\rho}\sim \rho$ in the IMF formula, Eq.~(\ref{eq:iMF}). 

By increasing the length $L$, the situation becomes more complicated. In general, $\tilde{\rho}$ first increases and saturates at $0.5$. Here the region on the left of the defect effectively acts as a (peculiar) reservoir and one can imagine that there is always a particle ready to be injected into the rightmost part. The duration of this transitory state increases with $L$, as there are more particles in the reservoir. After that, $\tilde{\rho}$ drops (low densities, Figures~\ref{fig:s6}a), corresponding to the shock passing the defect; then it converges to the homogeneous density in an oscillatory way, the oscillations representing the return of the diffusive shock (periodic boundaries).\\
\indent This behavior differs from the one observed in~\cite{motegi_exact_2012}, where the authors study a particular case of the relaxation of a TASEP with $\rho=0.5$, for which it is not possible to observe the oscillatory behavior.\\
\indent The study of the shock relaxation allows us to notice that, in large systems, the region closes again quicker than $\tilde{\rho}$ drops (or increases for high densities as in Fig.~\ref{fig:s6}b) or even relaxes to the TASEP density; this allows us to approximate $\tilde{\rho}$ with $0.5$ in the intermittent regime (see Sec.~\ref{sec::intermittent}). For instance, in Fig.~\ref{fig::FSMF}b we measured an average closing time $\langle t_f \rangle \sim 15$s for a system with $\rho=0.3$ and $L=500$. A system with the same features would need a time of $\sim$600s to escape the transitory state at $\tilde\rho \sim 0.5$, and longer times to eventually relax to the uniform TASEP density. Therefore, the IMF approach with $\tilde{\rho}=0.5$ gives a good approximation for large systems. Problems arise when $\tilde{\rho}$ drops (or increases) before the region closes, causing the observed non trivial current-density relationship for intermediate sizes.\\
 
\begin{figure}[h]
\begin{center}
\includegraphics[type=pdf,ext=.pdf,read=.pdf,width=\columnwidth]{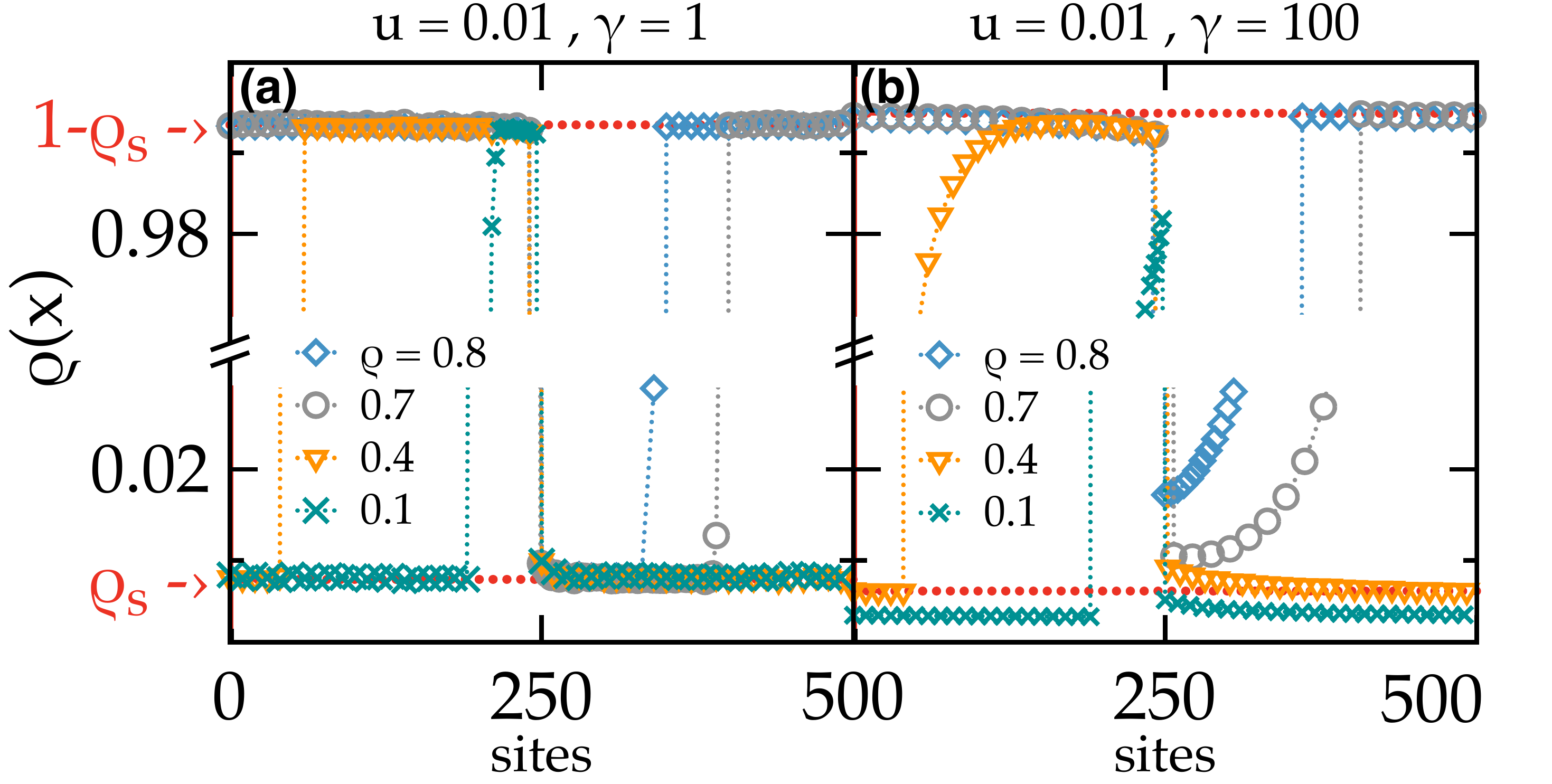}
\caption{(Color online) Density profiles for a lattice with $L=500, f=1s^{-1}, u=0.01s^{-1}$. (a) For $\gamma=1s^{-1}$ there is a clear HD-LD separation with a sharp front, in agreement with the FSMF (horizontal dotted lines). (b) In the presence of intermittence ($\gamma=100s^{-1}$) longer correlations are established and simulations deviate from the FSMF prediction.}
\label{profiles}
\end{center}
\end{figure}

Although in the other regimes the system also presents the ordinary finite-size effects of the TASEP, the severe finite-size effects in the intermittent regime have a different and rather counter-intuitive nature. They result from the transient relaxation of the density after an opening event and therefore present only in the intermittent regime. The occupancy of the defect site after an opening event indeed depends on the size of the system.\\  
\indent In this respect, intermittence and related finite-size effects have strong consequences on the stationary density profiles too. Whereas, similar to the static-defect case, outside the intermittent regime there is a sharp phase separation between the HD-LD profile before and after the dynamical defect (Fig.~\ref{profiles}a), the presence of intermittence induces relevant boundary effects modifying the density profiles (Fig.~\ref{profiles}b). \\
This is a signature of the transient relaxation of the HD-LD interface during the time $\langle t_f \rangle$. Such strong correlations are associated with the nonstationary dynamics of finite systems and are particularly evident at densities for which current reduction (or enhancement) occurs.

\section{Conclusion}
In this work we have introduced the concept of transport on a lattice in the presence of local interactions between particles and substrate, also referred to as \textit{dynamical defects}. 
This concept is key to understanding natural transport phenomena occurring on substrates with a fluctuating environment. It helps explain fundamental biological processes such as protein synthesis or intracellular traffic, where the local conformation of the substrate can deeply influence the characteristics of the flow of molecular motors. The obtained results, however, are general and therefore applicable to other transport processes, such as vehicular or human traffic, and synthetic molecular devices.\\
\indent The phenomenology is presented and studied by means of the totally asymmetric simple exclusion process, a prototypic model of transport in nonequilibrium physics. In this framework we have provided original mean-field arguments that allow the reproduction and the rationalization of the rich phenomenology of the model. We have thus discussed a novel dynamical regime, characterized by an intermittent current of particles and induced by the local interaction and competition between particle motion and the defect dynamics. Importantly, we have found that a particle-lattice interaction triggers severe finite-size effects that have a counterintuitive strong impact on transport. For different system densities, the small size of the lattice reduces or enhances the flow of particles, inducing an asymmetric current-density profile.\\
\indent Given the small size of biological substrates, the physics of the intermittent regime is highly relevant to the understanding of protein synthesis and motor protein transport.
Bursts of gene expression have been often reported~\cite{yu_probing_2006} and our results provide a physical mechanism to explain the contribution of the translation process to them. Moreover, intermittent behavior can strongly influence motor protein current fluctuations that can be measured in state of the art experiments~\cite{leduc_molecular_2012}. 
In the biological context of mRNA translation, our results on finite-size effects would correspond to an increase in protein production for small mRNA strands compared to longer strands; interestingly, highly expressed proteins constituents of ribosomes are short~\cite{planta_list_1998}.\\
\indent The model presented here can be extended by including, for example, larger interaction sites or different boundary conditions. However, the observed phenomenology does not change: if the defect is extended ($d>1$) the finite-size effects are even more pronounced than in the $d=1$ case~\cite{turci_preparation_2012}, and lattices with open boundaries, more common in practical applications than lattices with periodic boundary conditions, present the same characteristics, although edge effects can be relevant if the defect is moved close to the boundaries~\cite{turci_preparation_2012}.

\acknowledgments
We are grateful to I.~Stansfield for bringing to our attention this research topic, and to N.~Kern, I.~Neri and C.~A.~Brackley for valuable discussions. F.T. was supported by the French Ministry of Research, E.P. by CNRS and PHC no 19404QJ, L.C. by a SULSA studentship, M.C.R. by BBSRC (Grants No. BB/F00513/X1 and No. BB/G010722) and SULSA, and A.P. by the University of Montpellier 2 Scientific Council.

\appendix
\section{FSMF}
\label{FSMFapp}
If we denote the probability of being in any of the states of Table~\ref{table_states} by $P(x_{i})$, the time evolution of the probability vector $\vec{P}$ is governed by the master equation
\begin{equation}
\frac{\partial \vec{P}}{\partial t}=\mathbb{W}\vec{P}.
\end{equation}

Since we look for the steady state, we have $\mathbb{W}\vec{P}=\vec{0}$. Diagonalizing $\mathbb{W}$ one finds the eigenvector $\vec{P}=(p_{1},p_{2},\dots,p_6)$, where the components are functions of $\rho_s, u, f$ and $\gamma$.  Morover, the probability that the site $s$ is occupied is given by
\begin{equation}
\rho_{s}=p_{5}(\gamma, u,f,\rho_{s})+p_{6}(\gamma, u,f,\rho_{s}),
\label{finddensity}
\end{equation}
which yields a condition from which $\rho_{s}$ can be determined. 

Its analytical form, computed using {\sc Mathematica}\texttrademark , is given by

\begin{align}
\rho_{s}(u,f,\gamma)& =\frac{1}{6 \gamma (f+u)}\left\{2 f^2+5 f \gamma+4 f u+7 \gamma u+2 u^2+\right. \nonumber \\
&-(Q \sqrt{R})^{1/3}+ \left[-4 f^4-8 f^3 (\gamma+2 u)+\right.\nonumber\\
&+u^2 \left(-7 \gamma^2+2 \gamma u-4 u^2\right)+\nonumber\\
&-2 f u \left(5 \gamma^2+2 \gamma u+8 u^2\right)+ \nonumber\\
&\left.\left. -f^2 \left(7 \gamma^2+14 \gamma u+24 u^2\right)\right]/(Q \sqrt{R})^{1/3}\right\},
\end{align}
with
\begin{align}
R&=-\gamma^2 (f+u)^3 \left(16 f^7+32 f^6 (2 \gamma+3 u)+\right.\nonumber\\
&+9 u^3 \left(\gamma^2+\gamma u-2 u^2\right)^2+\nonumber\\
&+4 f^5 \left(22 \gamma^2+62 \gamma u+69 u^2\right)+\nonumber\\
&+ 4 f^4 \left(12 \gamma^3+51 \gamma^2 u+115 \gamma u^2+125 u^3\right)+\nonumber\\
&+ f u^2 \left(23 \gamma^4+14 \gamma^3 u-9 \gamma^2 u^2-8 \gamma u^3+196 u^4\right)+\nonumber\\
&+ f^2 u \left(23 \gamma^4+46 \gamma^3 u+127 \gamma^2 u^2+248 \gamma u^3+456 u^4\right)+\nonumber\\
&+\left.f^3 \left(9 \gamma^4+66 \gamma^3 u+225 \gamma^2 u^2+496 \gamma u^3+600 u^4\right)\right),
\end{align}
and
\begin{align}
Q&=-8 f^6-10 \gamma^3 u^3+39 \gamma^2 u^4+6 \gamma u^5-8 u^6+\nonumber\\
& -24 f^5 (\gamma+2 u)+3 \sqrt{3}-6 f^4 \left(\gamma^2+15 \gamma u+20 u^2\right)+\nonumber\\
&+3 f^2 u \left(2 \gamma^3+13 \gamma^2 u-20 \gamma u^2-40 u^3\right)+\nonumber\\
&-3 f u^2 \left(2 \gamma^3-27 \gamma^2 u+16 u^3\right)+\nonumber\\
&-f^3 \left(-10 \gamma^3+9 \gamma^2 u+120 \gamma u^2+160 u^3\right).
\end{align}
The current is then given by $J_{plateau}= \gamma \rho_{s} (1-\rho_{s})$ and compared to the outcome of numerical simulations in Fig.~\ref{folding-diag-composite}.

\end{document}